\documentclass[aps,prl,twocolumn,showpacs]{revtex4}
\usepackage{amsfonts}
\usepackage{mathrsfs}
\begin{document}
\title{Do gluons carry half of the nucleon momentum?}
\author{Xiang-Song Chen,$^{1,2,3,}$\footnote{cxs@scu.edu.cn}
Wei-Min Sun,$^3$ Xiao-Fu L\"{u},$^2$ Fan Wang,$^3$  and T.
Goldman$^{4,}$\footnote{t.goldman@post.harvard.edu}}
\affiliation{$^1$Department of Physics, Huazhong
University of Science and Technology, Wuhan 430074, China\\
$^2$Department of Physics,
Sichuan University, Chengdu 610064, China\\
$^3$Department of Physics, Nanjing University, CPNPC, Nanjing
210093, China\\
$^4$Theoretical Division, Los Alamos National Laboratory, Los
Alamos, NM 87545, USA}
\date{published 7 August 2009 in Phys. Rev. Lett. 103:062001}

\begin{abstract}
We examine the conventional picture that gluons carry about half of
the nucleon momentum in the asymptotic limit. We show that this
large fraction is due to an unsuitable definition of the gluon
momentum in an interacting theory. If defined in a gauge-invariant
and consistent way, the asymptotic gluon momentum fraction is
computed to be only about one fifth. This result suggests that the
asymptotic limit of the nucleon spin structure should also be
reexamined. A possible experimental test of our finding is discussed
in terms of novel parton distribution functions.

\pacs{11.10.Jj, 14.20.Dh, 12.38.-t, 11.15.-q}
\end{abstract}
\maketitle

A classic prediction of perturbative quantum chromodynamics
(QCD) is that about half of the nucleon momentum is carried by gluons
in the asymptotic limit~\cite{Poli74}. This renowned fraction has
become a deeply rooted picture in the minds of hadron physicists,
partially because it was supported~\cite{Sloa88}, indirectly, by
measurement of the quark momentum fraction in deeply inelastic
scattering (DIS) of leptons off nucleons (which we will
comment on later). Considered together with the nucleon
spin problem discovered later, however, the taken-for-granted
knowledge about the gluon contribution to the nucleon momentum
becomes questionable: Were the gluon momentum known perfectly, why
was severe difficulty (namely, with gauge invariance) encountered
with even a sound theoretical definition of the gluon orbital
angular momentum, which should be a direct extension of the gluon
momentum?

To better appreciate and clarify the issue, let us recall that the
conventional gluon momentum fraction is based on the following
decomposition of the total momentum operator in QCD:
\begin{equation}
\vec P_{\rm total} = \int d^3x \psi ^\dagger \frac 1i \vec D \psi
+\int d^3x \vec E \times \vec B \equiv \vec {\mathscr P}_q + \vec
{\mathscr P}_g \label{Poynting}.
\end{equation}
Here $\vec D=\vec \nabla -ig\vec A$ is the covariant derivative. (We
suppress all color indices and generators.) The quark and gluon
momentum fractions are defined through the nucleon-state expectation
values of the renormalized operators $\vec {\mathscr P}^R_q$ and
$\vec {\mathscr P}^R_g$. QCD prescribes the scale evolution of $\vec
{\mathscr P}^R_q$ and $\vec {\mathscr P}^R_g$ according to the
following well-known mixing matrix at leading order~\cite{Poli74}:
\begin{equation}
\gamma^{\mathscr P}=
 -\frac{\alpha_s}{4\pi} \left(
\begin{array}{cc}
-\frac 89 n_g& \frac 43 n_f\\
\frac 89 n_g& -\frac 43 n_f
\end{array}
\right), \label{matrix}
\end{equation}
which leads to the well-known asymptotic limit, $\vec {\mathscr
P}_g^R =\frac{2n_g}{2n_g+3n_f} \vec P_{\rm total}$, where $n_g=8$ is
the number of gluon fields and $n_f$ is the number of active quark
flavors. The most typical case of $n_f=5$ gives $\vec {\mathscr
P}^R_g \simeq \frac 12  \vec P_{\rm total}$, which says that gluons
carry about half momentum of a nucleon (or virtually any hadron) in
the asymptotic limit.

The doubtful point that we noted above is that the gluon momentum
density $\vec {\mathscr P}_g(\vec x)=\vec E \times \vec B$ (the
Poynting vector) in Eq. (\ref{Poynting}) cannot be used to construct
a gluon orbital angular momentum. In fact, in a pure Yang-Mills
theory (namely, without quarks) $\int d^3 x \vec x\times \vec
{\mathscr P}_g(\vec x)$ gives the conserved total angular momentum
of the gluon field, including both spin and orbital contributions:
\begin{equation}
\int d^3x \vec x\times (\vec E\times \vec B)=\int d^3x \vec E\times
\vec A +\int d^3 x \vec x \times E^i \vec \nabla A^i. \label{free}
\end{equation}

If we insist that a theoretically sound momentum density $\vec
P(\vec x)$ should match the orbital angular momentum in the standard
form $\int d^3 x \vec x\times \vec P(\vec x)$, then Eq. (\ref{free})
indicates that the gluon momentum density should be identified as
$\vec P_g(\vec x)=E^i \vec \nabla A^i$. It has often been argued
that there is no essential difference between $\vec P_g(\vec x)$ and
$\vec {\mathscr P}_g (\vec x)$ since they give the same integrated
gluon momentum. This equivalence, however, is limited to the pure
Yang-Mills case:
\begin{equation}
\vec E\times \vec B=E^i\vec \nabla A^i -\nabla^i(E^i\vec A)+(\vec
\mathcal{D}\cdot \vec E)\vec A. \label{identity}
\end{equation}
The second term is a total derivative and can be discarded in
integration. By the QCD equation of motion, the third term (where
$\vec \mathcal{D}=\vec \nabla -ig[\vec A,~]$ is the covariant
derivative for the adjoint representation) is zero in the absence of
quarks, but not zero otherwise! Hence we see that the quark-gluon
interaction renders $\int d^3x \vec {\mathscr P}_g(\vec x)$ and
$\int d^3x \vec P_g(\vec x)$ no longer equal. Then, naturally, we
may find a totally different picture of the nucleon momentum
partition if instead of $\int d^3x \vec {\mathscr P}_g(\vec x)$ we
define $\int d^3x \vec P_g(\vec x)$ as the gluon momentum. This
apparently interesting and important issue, however, has (to our
knowledge) never been pushed to actual calculations. The obstacle is
evident but not fully appreciated: that of gauge invariance, which
we elaborate in some detail in the following.

The density $\vec P_g(\vec x)$ is obviously gauge-dependent. The
gauge-invariance of its integration, $\int d^3x \vec P_g(\vec x)$,
is guaranteed in a pure Yang-Mills theory by the equality to the
gauge-invariant expression $\int d^3x \vec {\mathscr P}_g(\vec x)$.
When quarks are present, however, Eq. (\ref{identity}) says that
$\int d^3x \vec P_g(\vec x)$ loses its equality to $\int d^3x \vec
{\mathscr P}_g(\vec x)$, and thus loses simultaneously its property
of gauge invariance.

Being a different density, $\vec P_g(\vec x)$ does not give the same
angular momentum $\int d^3 x\vec x\times \vec P_g(\vec x)$ as $\int
d^3 x\vec x\times \vec {\mathscr P}_g(\vec x)$ even in a pure
Yang-Mills theory, and is always gauge-dependent. In fact, for
several decades it was stated in common textbooks that the
separation of spin and orbital angular momentum for a gauge field is
conceptually not possible, even for the Abelian case
\cite{textbook}. If applied to our momentum discussion, this seems to
dictate that the gluon momentum in the quark-gluon interacting case
is intrinsically not meaningful, since without a proper orbital
angular momentum definition we would have no means to match a
momentum expression. In a pure Yang-Mills theory, it is possible to ignore
 the gauge dependence of $\int d^3
x \vec x \times (E^i \vec \nabla A^i)$, and define the momentum
density as $E^i \vec \nabla A^i$, which, though gauge-dependent
itself, does give a gauge-invariant integrated momentum. However, as
we explained in the preceding paragraph, such a semi-justification
is lost in the quark-gluon interacting case.

It is quite tempting to agree with the common textbooks that in
gauge theories the basic physical notions like the gluon or photon
spin and orbital angular momentum are prohibited by gauge
invariance, because construction of these quantities necessarily
involves the gauge field $\vec A$. The gauge-invariance problem has
long prohibited a consistent investigation of the nucleon spin
structure \cite{Chen08}, and, fairly speaking, was simply overlooked
in the examination of the nucleon momentum.

Fortunately, a novel solution was achieved recently in
Ref.~\cite{Chen08}, which supplies a unified and gauge-invariant
treatment of the gluon momentum and angular momentum. The key to our
solution is to recognize that in the gauge coupling $\bar\psi
\gamma_\mu A^\mu\psi$, the gauge field $A^\mu$ plays a dual role: it
provides a physical coupling to the Dirac field $\psi$, as well as a
gauge freedom to compensate for the phase freedom of $\psi$. Our
idea for solving the gauge-invariance problem is to decompose this
dual role by seeking a unique separation $A^\mu= A^\mu_{\rm pure}+
A^\mu_{\rm phys}$, with $A^\mu_{\rm pure}$ a pure-gauge term,
transforming in the same manner as does the full $A^\mu$, and always
giving null field strength (i.e., $F^{\mu\nu}_{\rm pure}\equiv
\partial^\mu A^\nu_{\rm pure}-\partial^\nu A^\mu_{\rm
pure}+ig[A^\mu_{\rm pure},A^\nu_{\rm pure}]=0$), and $A^\mu_{\rm
phys}$, a physical term transforming in the same manner as does
$F^{\mu\nu}$. Namely, $A^\mu_{\rm phys}$ is gauge
invariant/covariant in Abelian/non-Abelian case, while $A^\mu_{\rm
pure}$ has the same gauge freedom as $A^\mu$ and can be used instead
of $A^\mu$ to construct a covariant derivative $ D^\mu_{\rm
pure}\equiv \partial^\mu+ig A^\mu_{\rm pure}$, acting on the
fundamental representation and $\mathcal{D}^\mu_{\rm pure}\equiv
\partial^\mu-ig[A^\mu_{\rm pure},~]$ acting on the adjoint
representation. The field separation is to express $A^\mu_{\rm
phys}$ and $A^\mu_{\rm pure}$ in terms of $A^\mu$, thus does not
interfere with the gauge condition or canonical quantization, which
manipulates the full $A^\mu$. In this way, the spin and orbital
angular momentum of the quark and gluon fields can all be
constructed gauge-invariantly at the density level:
\begin{eqnarray}
\vec J_{\rm total} &=&\int d^3x \vec E \times \vec A_{\rm phys}+
\int d^3x \vec x\times E^i \vec \mathcal{D}_{\rm pure} A_{\rm
phys}^i \nonumber\\
&&+ \int d^3 x \psi ^\dagger \frac 12 \vec \Sigma \psi + \int d^3x
\vec x \times\psi ^\dagger \frac 1i \vec D_{\rm pure} \psi .
\label{J}
\end{eqnarray}

Here we have improved over the expression in Ref.~\cite{Chen08} by
using the pure-gauge-covariant derivative for $\vec A_{\rm phys}$ as
well, which safely guarantees the gauge-invariance of $E^i \vec
\mathcal{D}_{\rm pure} A_{\rm phys}^i$. This is somewhat a natural
choice, since in QCD $A^\mu_{\rm phys}$ is gauge covariant instead
of invariant, thus as for $\psi$ or $F^{\mu\nu}$, a covariant
derivative should be applied. By doing so we are left with more
freedom in choosing a defining equation for $\vec A_{\rm phys}$
(e.g., $\vec {\mathcal D}_{\rm pure}\cdot \vec A_{\rm phys}=0$),
while in Ref.~\cite{Chen08} the equation had to include $[\vec
A_{\rm phys}, \vec E]=0$ so as to guarantee the gauge invariance of
$E^i \vec \nabla A_{\rm phys}^i$ with an ordinary derivative. We
have checked that all consistencies in Ref.~\cite{Chen08} are
maintained in the present formulation. Moreover, superior to $[\vec
A_{\rm phys}, \vec E]=0$, in the perturbative region $\vec {\mathcal
D}_{\rm pure}\cdot \vec A_{\rm phys}=0$ together with
$F^{\mu\nu}_{\rm pure}=0$ can give explicit series expression, in
powers of $g$, for $A^\mu_{\rm phys}$ (which has trivial boundary
behavior) and hence also $A^\mu_{\rm pure}=A^\mu-A^\mu_{\rm phys}$.
As was noted in Ref.~\cite{Chen08}, a proper definition of
$A^\mu_{\rm phys}$ for the non-Abelian theory is quite subtle, but
as we explain below, this subtlety does not affect the present
calculation.

From Eq. (\ref{J}), we can read out the corresponding
gauge-invariant quark and gluon momentum operators:
\begin{eqnarray}
\vec P_{\rm total} &=& \int d^3x \psi ^\dagger \frac 1i \vec D_{\rm
pure} \psi +\int d^3x E^i \vec \mathcal{D}_{\rm pure} A_{\rm phys}^i
\nonumber\\
 &\equiv& \vec P_q + \vec P_g \label{proper}.
\end{eqnarray}

It can be proven as in Ref. \cite{Chen08} that $\vec P_q + \vec P_g$
equals $\vec {\mathscr P}_q + \vec {\mathscr P}_g$ in Eq.
(\ref{Poynting}). Nevertheless, the individual terms are distinct in
the presence of quarks: $\vec {\mathscr P}_g$ differs from $\vec
P_g$ by a gauge-invariant quark-gluon interaction term:
\begin{equation}
\int d^3x \vec E\times\vec B=\int d^3x E^i \vec \mathcal{D}_{\rm
pure} A_{\rm phys}^i+ \int d^3x \psi^\dagger g\vec A_{\rm phys}\psi.
\label{diff}
\end{equation}

Equipped with the theoretically sound gluon momentum definition
which precisely matches an equally sound definition of the gluon
orbital angular momentum, we can now take up the previously
untouched calculation to reveal a possibly different picture of the
nucleon momentum partition. The calculation is straightforward and
parallels~\cite{Poli74} that based on the operators in Eq.
(\ref{Poynting}). Let us first display the mixing matrix for $\vec
P_q$ and $\vec P_g$:
\begin{equation}
\gamma^P=
 -\frac{\alpha_s}{4\pi} \left(
\begin{array}{cc}
-\frac 29 n_g& \frac 43 n_f\\
\frac 29 n_g& -\frac 43 n_f
\end{array}
\right), \label{master}
\end{equation}
which gives a different asymptotic limit for the renormalized gluon
momentum, $\vec P^R_g =\frac{n_g}{n_g+6n_f} \vec P_{\rm total}$. For
the typical case of $n_f=5$, this gives $\vec P^R_g\simeq \frac 15
\vec P_{\rm total}$, as compared to $\vec {\mathscr P}^R_g\simeq
\frac 12 \vec P_{\rm total}$.

Some technical details are worth mentioning. Because the total
momentum is conserved, the $2\times 2$ evolution matrix of the quark
and gluon momenta (in whatever definition) has only two independent
elements, as can be seen in Eqs. (\ref{matrix}) and (\ref{master}).
The easiest way of obtaining these two matrix elements is by
calculating the expectation value of the gluon momentum operator in
a quark state and the expectation value of the quark momentum
operator in a gluon state, which both start at order $\alpha_s$. By
doing so, we avoid the task of computing the quark and gluon
wavefunction renormalization if we consider only the leading-order
evolution matrix. This strategy is advantageous in obtaining Eq.
(\ref{matrix}), and becomes almost indispensable when obtaining Eq.
(\ref{master}). The reason is that computation of the gluon
wavefunction renormalization necessarily involves the non-Abelian
three-gluon vertex, while for 1-loop calculation of the gluon matrix
element in a quark state, the gluon field behaves like eight
independent Abelian fields. Unlike in the non-Abelian case, the
separation of an Abelian $A^\mu$ is unambiguous \cite{Chen08}. With
vanishing boundary values for perturbative calculations, $\vec
A_{\rm pure}$ is just the longitudinal field $\vec A_\parallel=\vec
\nabla \vec \nabla^{-2}(\vec \nabla\cdot\vec A)$, and $\vec A_{\rm
phys}$ is just the transverse field $\vec A_{\bot}=-\vec
\nabla\times \vec \nabla^{-2} (\vec \nabla \times \vec A)=\vec
A-\vec \nabla \vec \nabla^{-2}(\vec \nabla\cdot\vec A)$. ($A^0_{\rm
phys}$ and $A^0_{\rm pure}$ are not relevant here.) Therefore the
1-loop insertion of $\vec P_g$ in a quark state becomes simplest in
Coulomb gauge $\vec \nabla \cdot \vec A=0$, which gives $\vec A_{\rm
pure}=\vec 0$ and $\vec A_{\rm phys}=\vec A$. The calculation is
then a standard textbook exercise, and leads to the first column in
Eq. (\ref{master}). These two matrix elements are $\frac 14$ of
their counterparts in Eq. (\ref{matrix}). As to the 1-loop insertion
of $\vec P_q$ in a gluon state, we note that for a physical gluon,
$\int d^3x \psi^\dagger\vec A_{\rm phys}\psi$ vanishes identically
at 1-loop order. Then Eq. (\ref{diff}) means that $\vec P_q$ and
$\vec {\mathscr P}_q$ give the same 1-loop expectation value in a
physical gluon state, so the right-hand columns of Eqs.
(\ref{matrix}) and (\ref{master}) are the same.

Thus we see that if the gluon momentum is defined properly, its
contribution to the nucleon momentum is much smaller than
conventionally accepted. The difference can be traced to the fact
that the conventionally defined gluon momentum $\vec {\mathscr P}_g$
actually includes a quark-gluon interaction term, as indicated by
Eq. (\ref{diff}), while $\vec P_g$ is constructed solely in terms of
the physical degrees of freedom of the gluon field and is thus
justified as a proper gluon momentum.

The novel and distinct asymptotic limit of the nucleon momentum
partition we have presented naturally suggests that the asymptotic
limit of the nucleon spin partition be reexamined according to the
decomposition in Eq. (\ref{J}), where the gluon spin and orbital
angular momentum are constructed (as for $\vec P_g$) solely from the
physical gluon configuration. We expect that this investigation
would also reveal a distinct picture of the nucleon spin as compared
to that in Ref. \cite{Ji96}, which was based on a light-cone-gauge
formulation and thus included non-physical degrees of freedom in the
gluon angular momentum.

In the remainder of this paper, we discuss the delicate issue of
possible experimental measurement of the proper quark and gluon
momenta as defined by Eq. (\ref{proper}). To this end, it is very
illuminating to first recall how the interaction-involving momenta
in Eq. (\ref{Poynting}) are measured in hard processes. In fact, one
may ask a key question: Why are interaction-involving  quark and
gluon momenta measured, instead of the proper ones? The answer lies,
again, in gauge invariance.

Collision experiments do not measure the matrix elements of quark
and gluon momentum operators directly. These matrix elements are
extracted from the measured cross sections either via the operator
product expansion (OPE) analysis of DIS, or via their relation to
the quark and gluon parton distribution functions (PDFs) which can
be shown to factorize in DIS and other hard processes~\cite{Coll89}.
The reason for the conventional OPE approach to employ $\vec
{\mathscr P}_q(\vec x)$ and $\vec {\mathscr P}_g(\vec x)$ is
evident: Expansion of the products of the gauge-invariant quark
electric currents requires gauge-invariant operators. The naively
proper momentum operators $\psi^\dagger \frac 1i \vec \nabla \psi$
and $E^i\vec \nabla A^i$ are gauge dependent. Conventionally, the
only known way of accomplishing gauge invariance is by coupling a
gauge field to $\psi$, and by using $F^{\mu\nu}$ instead of $A^\mu$
in constructing physical quantities of the gauge-field. Either way,
the physical content of the original quantity is substantially
modified, because, as we explained earlier, a gauge field introduces
not only a gauge freedom but also a physical coupling. Moreover, the
restriction to $F^{\mu\nu}$ severely limits one's capability in
constructing certain quantities, e.g., the spin. Now that we have
decomposed the dual role of the gauge field by separating its
pure-gauge component from its physical component, we can take
exactly the needed part and discard the unwanted part for any
desired goal. Namely, solely to restore gauge invariance, we use
$A^\mu_{\rm pure}$ instead of the full $A^\mu$, while $A^\mu_{\rm
phys}$ is used instead of $F^{\mu\nu}$ whenever the gauge-field
canonical variable is unavoidable in constructing a physical
quantity. This is exactly how we succeed in constructing the
gauge-invariant and proper momentum, spin, and orbital angular
momentum of quarks and gluons in Eqs. (\ref{J}) and (\ref{proper}).
In principle, the gauge-invariant $\vec P_q(\vec x)$ and $\vec
P_g(\vec x)$ can be used instead of $\vec {\mathscr P}_q(\vec x)$
and $\vec {\mathscr P}_g(\vec x)$ in OPE analysis of DIS, so that
proper quark and gluon momenta can be measured.

It is less straightforward to see why the conventional factorization
approach also refers to the interaction-involving quark and gluon
momenta. In this approach, what one manipulates are the quark and
gluon PDFs, which factorize in cross sections for certain hard
processes~\cite{Coll89}, and can be integrated to give the quark and
gluon momenta by moment relations~\cite{Coll82}. Naively, one may
expect that a quark/gluon PDF should provide an unambiguous representation of
the quark/gluon property, but in conventional practice this is not
the case. The key point is how one actually constructs the quark and
gluon PDFs. Let us first look at the quark PDF in a target $A$
\cite{Coll82}:
\begin{eqnarray}
{\mathscr P}_{q/A}(\xi)=\int_{-\infty}^{\infty} \frac{dx^-}{4\pi}
e^{-i\xi P^+x^-}\langle \bar\psi(0,x^-,0_\bot) \gamma^+
\nonumber\\\times {\mathcal P}\exp\{ig\int_0^{x^-}dy^-
A^+(0,y^-,0_\bot)\} \psi(0)\rangle_A, \label{pdfq}
\end{eqnarray}
where a path-ordered (${\mathcal P}$) gauge link (Wilson line) is
inserted to achieve gauge invariance. It is exactly this gauge link
that makes the PDF defined by Eq. (\ref{pdfq}) an
interaction-involving one, since the gauge field brings not only a
gauge freedom but also a physical coupling to the quark field. The
interaction term is more evident in the moment relation:
$\int_{-\infty}^{\infty} d\xi \xi {\mathscr
P}_{q/A}(\xi)=\frac{1}{2(P^+)^2}\langle \bar \psi \gamma^+i
D^+\psi\rangle _A$. This is just the $+$ component of $\vec
{\mathscr P}_q(\vec x)$ in Eq. (\ref{Poynting}). Here the gauge
field in $D^+$ originates exactly from the gauge link in Eq.
(\ref{pdfq}). It is often taken for granted that gauge invariance
must be achieved at the price of a gauge coupling, and hence the
measurable momenta are always the interaction-involving ones in Eq.
(\ref{Poynting}).

But we do not really have to accept this confabulation for gauge
invariance. Solely to guarantee gauge invariance, the gauge link
can be constructed with the pure-gauge component $A^\mu_{\rm pure}$
instead of the full $A^\mu$, and thus we can define a proper and
gauge-invariant quark PDF:
\begin{eqnarray}
P_{q/A}(\xi)= \int_{-\infty}^{\infty} \frac{dx^-}{4\pi} e^{-i\xi
P^+x^-}\langle \bar\psi(0,x^-,0_\bot) \gamma^+ \nonumber\\
\times {\mathcal P}\exp\{ig\int_0^{x^-}dy^- A_{\rm
pure}^+(0,y^-,0_\bot)\} \psi(0) \rangle _A,
\end{eqnarray}
which integrates to give the desired proper quark momentum in Eq.
(\ref{proper}): $\int_{-\infty}^{\infty} d\xi \xi P
_{q/A}(\xi)=\frac{1}{2(P^+)^2}\langle \bar \psi \gamma^+ i D^+_{\rm
pure}\psi\rangle _A $.

Analogously, the conventional gluon PDF
\begin{eqnarray}
{\mathscr P}_{g/A}(\xi)= \int_{-\infty}^{\infty} \frac{dx^-}{2\pi\xi
P^+} e^{-i\xi P^+x^-}\langle F^{+\nu}(0,x^-,0_\bot) \nonumber
\\\times {\mathcal P}\exp\{ig\int_0^{x^-}dy^- A^+(0,y^-,0_\bot)\} F_\nu
^{~+}(0)\rangle _A
\end{eqnarray}
can be revised according to our strategy as
\begin{eqnarray}
P_{g/A}(\xi)= \int_{-\infty}^{\infty}  \frac{dx^-}{2\pi} e^{-i\xi
P^+x^-}\langle F^{+i}(0,x^-,0_\bot) \nonumber \\\times {\mathcal
P}\exp\{ig\int_0^{x^-}dy^- A_{\rm pure}^+(0,y^-,0_\bot)\} A^i_{\rm
phys}(0) \rangle _A,
\end{eqnarray}
where, besides the pure-gauge link, the physical component $A^i_{\rm
phys}$ is used instead of $F_\nu ^{~+}$ as the gauge-invariant
canonical variable. The second moments of ${\mathscr P}_{g/A}$ and
$P_{g/A}$ relate to the interaction-involving and proper gluon
momentum in Eqs. (\ref{Poynting}) and (\ref{proper}), respectively.

It can be expected that the proper quark and gluon PDFs, $P_{q/A}$
and $P_{g/A}$, should (though probably non-trivially) factorize in
the same processes to measure ${\mathscr P}_{q/A}$ and ${\mathscr
P}_{g/A}$, so that the proper quark and gluon momenta in Eq.
(\ref{proper}) can again be measured!

In closing, we remark that our approach is especially superior in
gauge-invariant construction of polarized and transverse-momentum
dependent PDFs with a clear particle number interpretation, and
off-forward PDFs which can be measured to infer the orbital angular
momenta in Eq. (\ref{J}), e.g., the polarized gluon PDF can be
defined gauge-invariantly as
\begin{eqnarray}
P_{\Delta g/A}(\xi)&=& \int_{-\infty}^{\infty}  \frac{dx^-}{2\pi}
e^{-i\xi P^+x^-}\langle F^{+i}(0,x^-,0_\bot) \nonumber
\\&&\times{\mathcal P}\exp\{ig\int_0^{x^-}dy^- A_{\rm
pure}^+(0,y^-,0_\bot)\}\nonumber \\
&&\times \epsilon_{ij+} A^j_{\rm phys}(0) \rangle _A,
\end{eqnarray}
with a first moment related to the gauge-invariant gluon spin in Eq.
(\ref{J}). Details will be reported elsewhere.

This work is supported by the China NSF under Grants No. 10875082,
No. 10475057, and No. 90503011, and the U.S. DOE under Contract No.
DE-AC52-06NA25396.


\begin{thebibliography}{99}
\bibitem{Poli74} H.D. Politzer, Phys. Rep. {\bf 14}, 129 (1974);
D.J. Gross and F. Wilczek, Phys. Rev. D {\bf 9}, 980 (1974); H.
Georgi and H.D. Politzer, {\it ibid.} {\bf 9}, 416 (1974).

\bibitem{Sloa88} T. Sloan, G. Smadja, and R. Voss, Phys.
Rep. {\bf 162}, 46 (1988).

\bibitem{textbook} See, e.g., J.M. Jauch and F. Rohrlich, {\it
The Theory of Photons and Electrons} (Springer-Verlag, Berlin 1976);
V.B. Berestetskii, E.M. Lifshitz, and L.P. Pitaevskii, {\it Quantum
Electrodynamics}, 2nd ed. (Pergamon, Oxford 1982).

\bibitem{Chen08} X.S. Chen, X.F. L\"u, W.M. Sun, F. Wang, and T.
Goldman, Phys. Rev. Lett. {\bf 100}, 232002 (2008).

\bibitem{Ji96} X. Ji, J. Tang, and P. Hoodbhoy, Phys. Rev. Lett.
{\bf 76}, 740 (1996).

\bibitem{Coll89} J.C. Collins, D.E. Soper, and G.
Sterman, in {\it Perturbative Quantum Chromodynamics}, ed. A.H.
Mueller (World Scientific, Sigapore 1989), p.1
[arXiv:hep-ph/0409313].

\bibitem{Coll82} J.C. Collins and D.E. Soper, Nucl. Phys. {\bf
B194}, 445 (1982).
\end{thebibliography}
\end{document}